\documentstyle[12pt]{article}
\begin{document}

\baselineskip=0.6cm

\noindent P.~N.~Lebedev Institute Preprint     \hfill
FIAN/TD/8--93\\ I.~E.~Tamm Theory Department       \hfill
\begin{flushright}{May 1993}\end{flushright}

\begin{center}

\vspace{0.5in}

{\Large\bf Superluminal Black Holes}

\bigskip

\vspace{0.3in}
{\large  D.~S.~Dolgov}\\
\medskip  {\it Department of Theoretical Physics} \\ {\it  P.~N.~Lebedev
Physical Institute} \\ {\it Leninsky prospect, 53, 117 924, Moscow,
Russia}$^{\dagger}$\\

\end{center}

\vspace{1.5cm}

\centerline{\bf ABSTRACT}
\begin{quotation}

The new solution of the Einstein equations in empty space
is presented.
The solution is constructed using Schwarzschild solution
but essentially differs from it. The basic properties of
the solution are: the existence of a horizon which is
a hyperboloid of
one sheet moving along its axis with superluminal velocity,
right signature of the metric outside the horizon and
Minkovsky-flatness of it at infinity outside the horizon.
There is also a discussion in the last chapter, including
comparing with recent astronomical observations.

\end{quotation}
\vfill
\noindent
$^{\dagger}$ E-mail address: ddolgov@td.fian.free.msk.su

\newpage
\pagestyle{plain}

\section{The metric}

Let $g_{\mu \nu} (\mu , \nu=0,1,2,3)$ be a Schwarzschild metric[2]:
\begin{equation}
dl^2=g_{\mu \nu} dx^\mu dx^\nu=-(1-\frac{2M}{r})dt^2+
     (1-\frac{2M}{r})^{-1} dr^2+r^2 (d\theta ^2+\sin ^2{\theta}d\phi ^2)
\end{equation}
where $r,\theta,\phi$ are connected with Cartesian coordinates via:
\begin{eqnarray}
&x&=r\sin{\theta}\cos{\phi}\nonumber\\
&y&=r\sin{\theta}\sin{\phi}\nonumber\\
&z&=r\cos{\theta}\nonumber
\end{eqnarray}

This metric is a well-known solution of the Einstein equations with
energy-momentum tensor~$T_{\mu \nu}$ equals to zero:
\begin{equation}
R_{\mu \nu}=0
\end{equation}
where $R_{\mu \nu}$ is a Ricci tensor.

This metric describes the geometry of a spherically symmetric black
hole and the external gravitational field of
a spherically symmetric mass~$M$.

Now let us write down the metric which corresponds to a moving
black hole (with some velocity, say~$v$). In order to do this
write down the Schwarzschild metric~(1) in Cartesian
coordinates~(~$t,x,y,z$):

\begin{eqnarray}
dl^2&=&-q dt^2+\nonumber \\
    &+&\frac{1}{q r^2}
      (x^2 dx^2+y^2 dy^2+z^2 dz^2+
      2xy dx dy+2xz dx dz+2yz dy dz)+\nonumber \\
    &+&\frac{1}{r^2}
      ((y^2+z^2)dx^2+(x^2+z^2)dy^2+(x^2+y^2)dz^2-\nonumber\\
    &-&2xy dx dy- 2xz dx dz-2yz dy dz)
\end{eqnarray}
where we denote $r\equiv\sqrt{x^2+y^2+z^2}$
and~$q\equiv 1-\frac{2M}{r}$.

Now passing to a moving along $x$-axis system of coordinates
and performing Lorentz transformation(boost along~$x$-axis):
\begin{eqnarray}
&t&^\prime=\frac{t+v x}{\sqrt{1-v^2}}\nonumber\\
&x&^\prime=\frac{x+v t}{\sqrt{1-v^2}}\nonumber\\
&y&^\prime=y\nonumber\\
&z&^\prime=z\nonumber
\end{eqnarray}
we obtain a metric which certainly satisfies the Einstein
equations~(2) (they are generally covariant)
and corresponds to a moving black
hole or a star:
\begin{eqnarray}
dl^2&=&-\frac{1}{(1-v^2)r^2}
      (q r^2-\frac{q^{-1}v^2}{1-v^2}(x-vt)^2-
       v^2(y^2+z^2))dt^2-\nonumber\\
     &-&2\frac{v}{(1-v^2)r^2}(-q r^2+
       \frac{q^{-1}}{1-v^2}(x-vt)^2+y^2+z^2) dt dx-\nonumber\\
     &-&2\frac{v(q^{-1}-1)}{1-v^2} \frac{(x-vt)y}{r^2} dt dy-
      2\frac{v(q^{-1}-1)}{1-v^2} \frac{(x-vt)z}{r^2} dt dz+\nonumber\\
     &+&\frac{1}{(1-v^2)r^2}(-v^2 q r^2+\frac{q^{-1}}{1-v^2}
       (x-vt)^2+y^2+z^2) dx^2+\nonumber\\
     &+&\frac{1}{r^2}
       (\frac{1}{1-v^2}(x-vt)^2+q^{-1}y^2+z^2) dy^2+
       \frac{1}{r^2}(\frac{1}{1-v^2}(x-vt)^2+y^2+
       q^{-1}z^2) dz^2+\nonumber\\
     &+&2\frac{q^{-1}-1}{1-v^2}\frac{(x-v t)y}{r^2}
       dx dy+2\frac{q^{-1}-1}{1-v^2}\frac{(x-v t)z}{r^2} dx dz+
       2(q^{-1}-1)\frac{yz}{r^2} dy dz
\end{eqnarray}
where we denote $r\equiv\sqrt{\frac{1}{1-v^2}(x-vt)^2
+y^2+z^2}$ and omit the primes.

Let us note four essential points about the obtained metric:
\begin{enumerate}
\item It is a solution of the Einstein equation at any value of~$|v|<1$;
\item It is asymptotically flat, i.e. becomes Minkovsky at infinity($r\to
\infty$);
\item It has Schwarzschild horizon (singularity) at $r=2M$;
\item It has the right signature ($-+++$) outside the horizon, i.e. when
\begin{equation}
\frac{1}{1-v^2}(x-vt)^2+y^2+z^2>(2M)^2
\end{equation}
\end{enumerate}

The crucial observation about the metric is that~(4) does not contain
the expressions~$\sqrt{1-v^2}$, although the Lorentz transformation
does (this is due to the fact that~(3) contains {\it only even}
powers of $x$ together with $dx$, $y$ together with $dy$ etc.).
This means the expression~(4) is symmetric bilinear and
real-valued form also at the values of~$|v|>1$
in the domain of definition (see below)
and we call it the metric.
Now we forget about the Lorentz transformations and the way we
obtained~(4). We know that for $-1<v<1$ the metric~(4)
is a solution of the Einstein equations which themselves do not depend on
parameter~$v$. Hence, for~$|v|>1$ the expression~(4) gives us the solution of
the Einstein equations as well (let us note that the components of the
metric~(4) are holomorphical functions on~$v$ except points~$v=\pm1$).

Now we should examine the basic properties of the obtained solution.
Let's fix some value of~$v$ ($|v|>1$).
Then the region of the space-time
where components of the metric take real values (i.e. the domain of
definition of the metric) is defined by:
\begin{equation}
\frac{1}{1-v^2}(x-v t)^2+y^2+z^2\ge0
\end{equation}
At any fixed value of $t$ this is 3-dimensional space except elliptic
cone, axis of which coincides with $x$-axis.
And as~$t$ grows
the picture moves along the $x$-axis with velocity~$v$~($|v|>1$).
This cone is the analogue of the point singularity in the
Schwarzschild metric.

We want to stress that the obtained metric at~$|v|>1$ can not be
transformed via coordinate transformations to usual Schwarzschild
metric (because the domain of definition of this metric is
not topologically equivalent to the domain of definition of
the Schwarzschild metric)
and, hence, gives us another solution of the Einstein
equations.

The obtained solution has a singularity,
corresponding to spherical Schwarzschild horizon in a black hole.
Now the singularity is a hypersurface defined by:
\begin{equation}
\frac{1}{1-v^2}(x-vt)^2+y^2+z^2=(2M)^2
\end{equation}
When $|v|>1$ this is a hyperboloid of one sheet
coaxial to $x$-axis and moving
along it with a velocity~$v$.

Next important feature of the metric is that it has the
right signature~($-+++$) in the region~(5). This region now is
an exterior of the hyperboloid at any fixed value of~$t$.

It is easy to see that the obtained solution becomes Minkovsky
at infinity at arbitrary direction in 3-dimensional space outside
the cones~(6)(at all these directions $q\to1$ at infinity). That is,
the solution is asymptotically flat at any direction in the domain
of definition~(6).

Also it is easy to show that, despite of~$|v|>1$, the horizon~(7)
is a time-like surface (i.e. normal vector is always space-like),
as in case of usual Schwarzschild metric.
Due to it there is possibility
for a particle to avoid a fall on the horizon with its velocity
less then that of light. In fact, we believe the particle in the
given geometry will move in the following way: it moves faster and
faster, approaching to the horizon and simultaneously sliding
along it and its velocity approaches to that of light. The work
of finding exact trajectories of the particles is now in progress.

The solution~(4) can be rewritten in a simple form, using the
new coordinates-($t^\prime,\rho,\chi,\psi$) defined as follows:
\begin{eqnarray}
&t&^\prime=\frac{t-vx}{\sqrt{v^2-1}} \nonumber\\
&\rho& \sinh{\chi}=\frac{x-vt}{\sqrt{v^2-1}}\nonumber\\
&\rho& \cosh{\chi} \sin{\psi}=y \nonumber\\
&\rho& \cosh{\chi} \cos{\psi}=z
\end{eqnarray}
The metric~(4) becomes:
\begin{equation}
dl^2=(1-\frac{2M}{\rho}){dt^\prime}^2+(1-\frac{2M}{\rho})^{-1}d\rho^2
+\rho^2(-d\chi^2+\sinh{\chi}^2 d\psi^2)
\end{equation}

Direct calculations show that this metric satisfies the
Einstein equations~(2).

But these coordinates are unphysical because of the signs in the
expression above. For example,~$t^\prime$ cannot be considered
as a time variable since the term containing~${dt^\prime}^2$
has the wrong sign. But let us stress that the metric itself
is "good" in the former coordinates~($t,x,y,z$), which can
be considered as physical(Cartesian) coordinates.

In fact, the solution~(4) is not the parametrized family of
solutions (with parameter~$v$) of
the Einstein equations but only one solution and all others can
be obtained via Lorentz transformation of the given one. We may
chose this standard solution in order to factor~$\frac{1}{1-v^2}$
equals to~$-1$, i.e.~$v=\sqrt{2}$. In that case, for obtaining
the solution~(4) at some value~$v_0$ (including~$v_0=\infty$)
one performs Lorentz transformations(boost along~$x$-axis) with
\[
v_L=\frac{\sqrt{2}-v_0}{\sqrt{2}v_0-1}
\]
It is possible to chose as the standard solution the solution~(4)
with~$v=\infty$. This case will be examined in the next chapter.

\section{Limit $v\to\infty$}

Now let's see what solution do we obtain when~$v$ approaches to
infinity. Substituting~$v~\to~\infty$ into~(4) we find:
\begin{eqnarray}
dl^2&=&-\frac{1}{r^2}(-q^{-1} t^2+y^2+z^2)dt^2-
        2(q^{-1}-1)\frac{t y}{r^2} dt dy-
        2(q^{-1}-1)\frac{t z}{r^2} dt dz+q dx^2+\nonumber\\
     &+&\frac{1}{r^2}(-t^2+q^{-1} y^2+z^2)dy^2+
              \frac{1}{r^2}(-t^2+y^2+q^{-1} z^2)dz^2+
             2(q^{-1}-1)\frac{y z}{r^2}dy dz
\end{eqnarray}
where $r\equiv\sqrt{y^2+z^2-t^2}$ and~$q\equiv1-\frac{2M}{r}$.

There is a singularity in this metric, corresponding to the
Schwarzschild horizon, when $q=0$, i.e. when
\begin{equation}
y^2+z^2-t^2=(2M)^2
\end{equation}
The sections of this hypersurface by the hyperplanes~$t=$const
are cylinders coaxial with~$x$-axis and their radius depend on~$t$
as follows:
\[
R(t)=\sqrt{(2M)^2+t^2}
\]

The arbitrary solution~(4) at some particular value~$v$
can be obtained from this solution
via boost along~$x$-axis with~$v_L=1/v$.

Calculation shows that in this case($v=\infty$) the light, emitted
at the horizon~(7) will not leave the horizon (as in usual black
hole). Since any other solution at any value of~$|v|>1$ can be
obtained from this one via Lorentz transformation, the situation
is the same for any~$v$ (at arbitrary~$v$ the horizon is a one-
sheet hyperboloid moving along its axis with a velocity~$v$.
Thus, we may say that the horizon (moving with a velocity greater
then that of light) has a definite physical properties and it is
not an imaginable surface.

\section{Discussion, Speculations and Conclusion}

The obtained solution was called "new", because it was not
found in the book~[1].

The basic question arising about the obtained solution is its
physical meaning. It is well-known that the Einstein equations
have nonphysical solutions even if we consider solutions, which
become Minkovsky-flat at infinity. The main problem arising
is the causality principle violation. The presented solution
has the right signature in the exterior of the horizon
(hyperboloid) and that's why the particle in the gravitational
field, corresponding to the solution, will move slower then
light. Also it seems that the particle cannot fall under the
horizon. These are arguments that the causality principle
is not violated in this case.

It is possible to make an essential refinement of the solution
if we map the domain of definition of the metric (i.e. exterior
of elliptic cone at fixed~$t$) to the whole 4-dimensional space
except 2-dimensional plane (which sections by hyperplanes~$t=const$
are the lines, coinciding with hyperboloid axis). These
are isotopic, so this mapping(isotopy) is possible. This mapping is
ambiguous and it can be chosen in order to transform the horizon
from the hyperboloid to a "spindle" continuing to infinity in the
both directions of its edges. Such a solution seems to be more
physically valuable. Yet it is not clear, is it possible to
arrange the mapping in order to make the metric Minkovsky-flat at
infinity at any direction in the resulting space (at least in
some range of directions this can be done). This point is very
important because if it is so, one can consider the "spindle"-
solution as possibly existing in the Universe.

Thus the  General Theory of Relativity(GTR) together with the
causality principle do not forbid a movement with a
superluminal velocity. Probably, the important condition
of such a movement is the presence of a horizon which
itself moves faster then light, through
which it is impossible for a particle to pass in a finite
time and, hence, it is impossible to transfer an information
(even just increasing a mass of a superluminal object) with
a superluminal velocity. Since the existence of horizons
is one of predictions of the GTR, finding out in the Universe
superluminal
objects would be else one argument in favor of it.

Let us suppose that there are superluminal objects
(relatively local) in the Universe and we can see their
traces (or maybe radiation of the object). How we would
see their movement? The answer is: we would see first
appearing of two objects at some place in the Universe
and second scattering of them into opposite directions.
There is an illusion of a backward
movement when an object {\it approaches \/} an
observer. And when an object moves {\it off \/}
an observer there is no such illusion. Hence,
an observer see a pair of resembling objects
scattering into opposite directions.

The presented solution of the Einstein equation
gives some theoretical possibility of the existence
of such superluminal object. Is there an astronomical
evidence of such objects (since the pairs of
scattering objects should be quite noticeable)?

The answer is yes.
One of the first observations of that kind was made by
Arp in 1967 [3]. He observed, when studying peculiar galaxies,
that very often a peculiar galaxy lies on the line, joining
two radio-sources between them.
Such examples are statistically
significant, so one may say that these are physically
connected and the radio sources was ejected from the
galaxy. But, in this case, there arise some problems:
first, usually one thinks that radio sources are farther
then associated peculiar galaxy and, second, one
should explain the locality of radio sources.
In addition these objects often have
close values of their parameters (such
as redshifts and magnitude)
The assumption that these pairs are superluminal objects
clearly explains these observations.
Let me note that except these pairs there are
also many other examples of pairing and
lining
quite local objects in Universe (for example, pairing
of quasars with large separation on the sky[7] and lining
of superclusters of galaxies).

Probably the assumption of existence in the Universe
superluminal objects also can explain
some strange phenomena recently discovered by
astronomers[3,4,5,6].

\section*{Acknowledgments}

I would like to thank Prof.V.Ya.Fainberg, Prof.I.V.Tyutin,
Prof.B.L.Voronov and M.Zelnikov for very useful discussions.

\section*{References}

1. {\it Exact Solutions of the Einstein field equations \/}
(ed.E.Schmutzer, Berlin,~1980)

\noindent
2. Misner C., Thorne K., Wheeler J.{\it Gravitation\/}(Freeman,~1973)

\noindent
3. Arp H. {\it Astrophys.J.\/}{\bf 148}(1967) 321

\noindent
4. Arp H. {\it Quasars, redshifts and controversies\/}(Berkley,~1987)

\noindent
5. Burbidge G. et all {\it Astrophyis.J.Suppl.\/}{\bf 74}(1990)n.3

\noindent
6. Arp H., Burbidge G., Hoyle F., Narlikar~J.V., Wickrama~N.C.
{\it Nature\/}{\bf 346}(1990) 807

\noindent
7. Arp H., Bi H.G. Max Plank Institute preprint 516, 1990

\end{document}